\PassOptionsToPackage{merge}{natbib}
\documentclass[epj]{webofc}
\usepackage[varg]{txfonts}   % Web of Conferences font
\wocname{EPJ Web of Conferences}
\woctitle{ICNFP 2016}
\usepackage{mathtools}

\newcommand{\bi}{\bigskip}
\newcommand{\no}{\noindent}
\newcommand{\be}{\begin{equation}}
	\newcommand{\ee}{\end{equation}}
\newcommand{\bea}{\begin{eqnarray}}
	\newcommand{\eea}{\end{eqnarray}}

\newcommand{\sli}{\sum\limits}

\newcommand{\il}{\int\limits}

\def\RR{{\mathbb{R}}}

\newcommand{\vB}{\vec{B}}

\newcommand{\vA}{\vec{A}}
\newcommand{\vx}{\vec{x}}
\newcommand{\vy}{\vec{y}}
\newcommand{\vD}{\vec{D}}
\newcommand{\vk}{\vec{k}}
\newcommand{\vp}{\vec{p}}
\newcommand{\vq}{\vec{q}}

\DeclareMathOperator{\Tr}{Tr}

\newcommand*{\dbar}{\mathop{}\!\mathrm{d}\mkern1.5mu\mathllap{\mathchar'26}\mkern-.5mu}

\renewcommand{\vec}[1]{\boldsymbol{#1}}

\begin{document}
\selectlanguage{english}
\title{Hamiltonian approach to QCD at finite temperature\footnote{Invited talk given by H.~Reinhardt at ICNFP2018, Kolymbari, Greece.}}
%
% subtitle (optional, strongly discouraged)
%
%%%\subtitle{Do you have a subtitle?\\ If so, write it here}

\author{H. Reinhardt\inst{1}\fnsep\thanks{\email{hugo.reinhardt@uni-tuebingen.de}} \and
	D.~Campagnari\inst{1}\and M.~Quandt\inst{1}
 }

\institute{Universit\"at T\"ubingen, Institut f\"ur Theoretische Physik, Auf der Morgenstelle 14, 72076 T\"ubingen, Germany}

\abstract{%
  A novel approach to the Hamiltonian formulation of quantum field theory at finite temperature is presented. The temperature is introduced by compactification of a spatial dimension. The whole finite-temperature theory is encoded in the ground state on the spatial manifold $S^1 (L) \times \RR^2$ where $L$ is the length of the compactified dimension which defines the inverse temperature. The approach which is then applied to the Hamiltonian formulation of QCD in Coulomb gauge to study the chiral phase transition at finite temperatures.
}
\maketitle
\section{Introduction}
\label{intro}
The understanding 
of the QCD phase diagram is subject of intensive studies both experimentally and theoretically. On the lattice we have access to the finite-temperature behaviour of hadronic matter. 
The lattice methods developed so far fail, however, at large chemical potential due to the notorious sign problem. Therefore non-perturbative continuum methods are needed. In the recent 10 to 15 years much efforts were undertaken to develop such non-perturbative continuum approaches. From the conceptual point of view we can distinguish three basic approaches:
\begin{enumerate}
	\item Dyson--Schwinger equations (DSE) \cite{Fischer2006,Alkofer2001, Binosi2009,Watson2006,Watson2007,Watson2008},
	\item Functional renormalization group flow equations \cite{Pawlowski2007,Gies2012},
	\item Variational methods both in covariant form \cite{Quandt2013,Quandt2015} and in the Hamiltonian formulation \cite{Feuchter2004,Feuchter2005}.
\end{enumerate}
There are also semiphenomenological approaches using a massive gluon propagator \cite{Reinosa:2014zta,*RSTW2016} or the Gribov--Zwanziger action \cite{Gribov1978,*Zwanziger:1988jt,*Zwanziger:1989mf,*Zwanziger:1992qr,Canfora2015}.
In this talk I will discuss a variational Hamiltonian approach to QCD at finite temperature \cite{Feuchter2004,Feuchter2005}.
\bi

\no
At finite temperature the central quantity is the partition function and in the Hamiltonian approach this quantity is calculated as the trace of the grand canonical density operator $\exp (- H/T)$:
\begin{eqnarray}
	\label{161-1}
	Z (L) = \Tr \mathrm{e}^{- H/T} \, ,
\end{eqnarray}
where the chemical potential was set to zero for simplicity. 
For a quantum field theory the Hamiltonian $H$ contains interactions, which makes the treatment of the density operator very cumbersome and one usually resorts to a quasi-particle approximation for the Hamiltonian $H$ in the exponent so that Wick's theorem can be used. 
There is, however, a much more efficient way to study quantum field theory at finite temperatures  in the Hamiltonian formulation without explicitly calculating the whole partition function and thus without the need to introduce additional approximations to the density operator. This new approach \cite{Reinhardt:2016xci} is based on the compactification of a spatial dimension. 
The finite temperature theory is fully encoded in the vacuum state on the partially compactified manifold $S^1 (L) \times \RR^2$.
In my talk I will use this novel approach to study QCD at finite temperature. I will first explain this approach and then apply it to the Hamilton formulation of QCD in Coulomb gauge to study the chiral phase transition. For this purpose I will first present the basic ingredients to the variational Hamiltonian approach to QCD in Coulomb gauge and summarize some essential zero-temperature results. 
Within this approach I will then study the quark sector at finite temperature where the focus is put on the chiral phase transition.

\section{Finite temperature from compactification of a spatial dimension}\label{sect2}
\label{sec-2}
The current standard approach to finite-temperature quantum field theory is to express the partition function (\ref{161-1}) by the functional integral
\begin{equation}
	\label{G2}
	Z (L) = \il_{b.c.} D (A, \psi) \exp \left[
	- \il^{L/2}_{- L/2} d x^0 \int d^3 x L_E (A, \psi)
	\right] \, ,
\end{equation}
where the bosonic (fermionic) fields satisfy (anti-)periodic boundary conditions in the Euclidean time 
\begin{eqnarray}
	\label{G3}
	A (x^0 = L/2) & =& A (x^0 = - L/2) \, , \nonumber\\
	\psi (x^0 = L/2) & = & - \psi (x^0 = - L/2) \, .
\end{eqnarray}
These boundary conditions compactify the Euclidean time axis to a circle with circumference $L$ so that the space-time manifold of the finite-temperature theory is $S^1 (L) \times \RR^3$. 
\bi

\no
In a relativistic invariant quantum field theory the Euclidean Lagrange density $L_E (A, \psi)$ is O(4) invariant. One can exploit this invariance to rotate the time axis onto a spatial axis and consequently one of the spatial axes onto the time axis, e.g. 
\begin{eqnarray}
	\label{G4}
	x^0 \to x^3 \, , \qquad A^0 \to A^3 \, , \qquad \gamma^0 \to \gamma^3 \, , \nonumber\\
	x^1 \to x^0 \, , \qquad A^1 \to A^0 \, , \qquad \gamma^1 \to \gamma^0 \, .
\end{eqnarray}
A O(4) rotation transforms all Lorentz vectors in the same way so, together with the space time coordinates $x^\mu$, we have also to rotate the gauge field $A^\mu$ and the Dirac matrices $\gamma^\mu$. As a consequence of such an O(4) rotation the temporal boundary conditions (\ref{G3}) become spatial boundary conditions 
\begin{eqnarray}
	\label{G5}
	A (x^3 = L/2) & = & A (x^3 = - L/2) \, , \nonumber\\
	\psi (x^3 = L/2) & = & - \psi (x^3 = - L/2) \, .
\end{eqnarray}
These boundary conditions compactify the 3-axis to a circle $S^1 (L)$ so that the spatial manifold is now $S^1 (L) \times \RR^2$. 
\bi

\no
Reversing now the steps which lead from the partition function in the Hamiltonian form (\ref{161-1}) to the functional integral form (\ref{G2}), which implies the canonical quantization on the spatial manifold $S^1 (L) \times \RR^2$, we obtain the representation \cite{Reinhardt:2016xci}
\begin{equation}
	\label{G6}
	Z (L) = \lim\limits_{l \to \infty} \Tr \mathrm{e}^{- l H (L)} = \lim\limits_{l \to \infty} \sli_n \mathrm{e}^{- l E_n (L)} \, ,
\end{equation}
where $l$ denotes the length of the uncompactified spatial dimensions which, of course, is infinity. Furthermore, $H (L)$ is the Hamiltonian on the spatial manifold $S^1 (L)  \times \RR^2$ and $E_n (L)$ are its eigenvalues. 
\bi

\no
Note that in the partition function (\ref{G6}) the inverse temperature $L$ no longer multiplies the Hamiltonian. Instead the latter is defined on the spatial manifold $S^1 (L) \times \RR^2$ and thus temperature dependent. Since $l \to \infty$ from the sum over the energy eigenstates only the ground state survives
\begin{equation}
	\label{G7}
	Z (L) = \lim\limits_{l \to \infty} \exp (- l E_0 (L)) \, .
\end{equation}
Therefore the whole partition function is given by the vacuum energy $E_0 (L)$ on the spatial manifold $S^1 (L) \times \RR^2$. In this way we have reduced the calculation of the partition function to the evaluation of the ground state energy $E_0 (L)$ on the spatial manifold $S^1 (L) \times \RR^2$. The only assumption made in the derivation of Eq.~(\ref{G7}) was the O(4) invariance of the Euclidean Lagrange density, which holds for any relativistically invariant theory. It does, however, not hold for a non-relativistic many-body system.
\bi

\no
From the partition function (\ref{G7}) we can derive the desired thermodynamic quantities in the standard fashion by taking derivatives. For the pressure $P$ and energy density $\varepsilon$ we obtain 
\begin{eqnarray}
	\label{G8}
	P & = & - \partial [Ve (L)] / \partial V \, , \qquad V = l^3 \, , \\
	\label{G9}
	\varepsilon & = & \partial [L e (L)] / \partial L - \mu \partial e (L) / \partial \mu \, ,
\end{eqnarray}
where $e (L)$ is defined by separating the ``spatial'' volume $L l^2$ from the ground state energy $E_0 (L) = L l^2 e (L)$ and is referred to as pseudo-energy density. Finally in the presence of a finite chemical potential for the fermions the Dirac Hamiltonian receives an extra term \cite{Reinhardt:2016xci}
\be
\label{G10}
h = \vec{\alpha} \vp + \beta m \to h (L) = h + i \mu \alpha^3 \, ,
\ee
where $\alpha^3$ is the Dirac matrix corresponding to the compactified space dimension.
\bi

\no
As we have seen above the whole finite-temperature quantum field theory can be entirely extracted from the vacuum state on the spatial manifold $S^1 (L) \times \RR^2$ and there is no need to explicitly calculate the partition function. In particular, no excited states have to be determined. However, one pays a price: on the spatial manifold $S^1 (L) \times \RR^2$ the usual O(3) invariance is, of course, broken to O(2), which will complicate the explicit calculations, e.g. the integral over a function in momentum space in $\RR^3$, $\int d^3 p f (\vp)$, is replaced on $S^1 (L) \times \RR^2$ by 
\be
\label{G11}
\il_L d^3 p f (\vp) := \il d^2 p_\perp \frac{2 \pi}{L} \sli^\infty_{n = - \infty} f (\vp_\perp , \omega_n) \, ,
\ee
where
\be
\label{G12}
\omega_n = \left\{
\begin{array}{ccl}
	\frac{2 \pi n}{L} & , & \text{bosons} : n_F = 0 \\[3mm]
	\frac{(2 n + 1) \pi}{L} & , & \text{fermions} : n_F = 1
\end{array}
\right.
\ee
are the Matsubara frequencies. For O(3)-invariant observables $f (\vp) = f (|\vp|)$ the 3-dimensional integral $\int d^3 p \, f (\vp)$ reduces to a 1-dimensional integral over the modulus of the momentum with integrals over the angles being trivial while on the spatial manifold $S^1 (L) \times \RR^2$ we have in addition a summation over the Matsubara frequencies and instead of the single function $f (|\vp|)$  we have an (in principle infinite) set of functions $f (|\vp_\perp|, \omega_n)$. Fortunately, as the temperature increases fewer and fewer Matsubara frequencies have to be included and in the high-temperature limit only the lowest Matsubara frequency $\omega_{n = 0}$ survives. For small temperatures a huge number of Matsubara frequencies have to be included. In that case, however, it is more convenient to perform a Poisson resummation 
\begin{equation}
	\label{G13}
	\frac{1}{2 \pi} \sli^\infty_{k = - \infty}
	\mathrm{e}^{ikx} = \sli^\infty_{n = - \infty} \delta (x - 2 \pi n)
\end{equation} 
by means of which the integral Eq.~(\ref{G11}) is converted to 
\begin{align}
	\label{G14}
	\il_L d^3 p f (\vp) = \int d^2 p_\perp \int d p_3 f (\vp_\perp, p_3) \sli^\infty_{k = - \infty} (-)^{n_F k} \exp (i k L p_3) \, .
\end{align}
The momentum integral is here the same as on $\RR^3$ (i.e.~as in the zero-temperature case). However, we have here in addition a summation over an index $k$ of the oscillating function $\exp (i k L p_3)$. Note that now the temperature enters explicitly only through the oscillating function $\exp (i k L p_3)$. The term with $k = 0$ is independent of the temperature and obviously agrees with the corresponding zero-temperature expression. Here an advantage of the Poisson resummed form Eq.~(\ref{G14}) becomes apparent: The zero-temperature vacuum contribution to an observable, which is usually divergent, can be easily extracted. In fact, in the zero-temperature limit $L \to \infty$ only the $k = 0$ term survives while the terms with $k \neq 0$ become rapidly oscillating functions with give zero contribution to any momentum integral. As the temperature increases more and more terms $k  \neq 0$ have to be included. For the study of a phase transition at finite temperature it is convenient to do the calculations from both ends starting at zero temperature with the Poisson resummed form (\ref{G14}) and at  high temperature with the Matsubara sum (\ref{G11}), and extend the calculations to an overlap regime containing the critical temperature where both approaches are applicable with a moderate number of terms included.
\bi

\no
Let us illustrate this novel approach to Hamiltonian quantum field theory at finite temperatures by considering a gas of non-interacting massive bosons and fermions with a single particle energy
\begin{equation}
	\label{G15}
	\omega (\vp) = \sqrt{m^2 + \vp^2} \, ,
\end{equation}
In the usual formulation of finite-temperature quantum field theory the pressure of the Bose gas is given by
\begin{equation}
	\label{G16}
	P = \frac{2}{3} \int \dbar^3 p \, \frac{\vp^2}{\omega (\vp)} \, n (\vp) , \qquad n(\vp) = \left( \mathrm{e}^{L \omega (\vp)} - 1 \right)^{- 1} ,
\end{equation}
where $n(\vp)$ are the finite-temperature Bose occupation numbers. In the novel approach presented above one finds for the pressure from Eq.~(\ref{G8}) $P = - e (L)$, and the pseudo-energy density $e (L)$ for a free gas of bosons is given by their ground state energy density on $S^1 (L) \times \RR^2$
\begin{equation}
	\label{G17}
	e (L) = \frac{1}{2} \il_L \dbar^3 p \, \omega (\vp) = \frac{1}{2} \int \dbar^2 p_\perp \frac{1}{L} \sli^\infty_{n = - \infty} \sqrt{\vp^2_\perp + \omega^2_n} \, .
\end{equation}
The two expressions given by Eq.~(\ref{G16}) and (\ref{G17}) are certainly not identical. In fact, the last expression (\ref{G17}) is even ill defined. To make it well defined we represent this square root by a proper time integral
\begin{equation}
	\label{G18}
	A = \frac{1}{\Gamma (- 1/2)} \lim\limits_{\Lambda \to \infty} \il^\infty_{1/\Lambda^2} d \tau \tau^{- 3/2} \mathrm{e}^{- \tau A^2} \, .
\end{equation}
Then the momentum integration can be carried out. Using the Poisson resummed form (\ref{G14}) we obtain 
\begin{equation}
	\label{G19}
	e (L) = - \frac{1}{2 \pi^2} \sli^\infty_{n = - \infty} \left(
	\frac{m}{n L}
	\right)^2 K_{- 2} (n L m) \, ,
\end{equation}
where $K_{- 2} (x) = K_2 (x)$ is the modified Bessel function. The term with $n = 0$ represents the energy density of the Bose gas on $\RR^3$. This is the usual vacuum $(T = 0)$ contribution which is divergent and which has to be omitted as usual. Taking in the remaining sum the limit $m \to 0$ we find for the pressure of the massless Bose gas 
\begin{equation}
	\label{G20}
	P = \frac{\xi (4)}{\pi^2} T^4 = \frac{\pi^2}{90} T^4 \, ,
\end{equation}
which is the correct expression.
\bi

\no
For a non-interacting Fermi gas the pseudo-energy density is given by the energy of the Dirac sea on $S^1 (L) \times \RR^2$. For fermions with the energy $\omega (\vp)$ (\ref{G15}) (on $\RR^3$) at a finite chemical potential $\mu$ we find by using Eq.~(\ref{G10}) 
\begin{equation}
	\label{G21}
	e (L) = - 2 \il_L \dbar^3 p \, \omega (\vp_\perp, p_3 + i \mu) \, ,
\end{equation}
where the factor of 2 comes from the two spin degrees of freedom.
Repeating the manipulations as carried out in the bosonic case one finds 
\begin{equation}
	\label{G22}
	e (L) = - \frac{2}{\pi^2} \sli^\infty_{n = - \infty} \cos \left[
	n L \left(
	\frac{\pi}{L} - i \mu
	\right)
	\right]
	\left(
	\frac{m}{n L}
	\right)^2 K_{- 2} (n L m) \, .
\end{equation}
Again we have to omit the $n = 0$ term, which represents the vacuum $T = 0$ contribution to the energy density. Even after removing the $n = 0$ term, the remaining sum is ill defined. To make it well defined we continue the chemical potential to pure imaginary values, $i \mu \to x$, and use 
\be
\label{G23}
\sli^\infty_{n = 1} (-)^n \frac{\cos (n x)}{n^4} = \frac{1}{48} \left[
	- \frac{7}{15} \pi^4 + 2 \pi^2 x^2 - x^4
	\right] \, .
	\ee
	Continuing the obtained result back to real chemical potentials one finds for massless fermions
	\be
	\label{G24}
	P = \frac{1}{12 \pi^2} \left[
	\frac{7}{15} \pi^4 T^4 + 2 \pi^2 T^2 \mu^2 + \mu^4
	\right] \, ,
	\ee
	which is the correct expression.
	\bi
	
	\no
		In the following I apply the approach to finite-temperature Hamiltonian quantum field theory \cite{Reinhardt:2016xci} summarized above to QCD in Coulomb gauge, where I will focus on the chiral phase transition at finite temperatures. 

	\section{Hamiltonian approach to QCD in Coulomb gauge}
	\subsection{Yang-Mills sector}\label{sect3.1}
	
	The Hamiltonian approach usually starts from the Weyl gauge $A_0 = 0$ which fixes the gauge up to time-independent gauge transformations. The latter can be conveniently fixed by using the Coulomb gauge $\vec{\partial} \vA = 0$, which has the advantage that Gauss' law can be explicitly resolved resulting in the gauge fixed Hamiltonian \cite{Christ1980}
	\be
	\label{G25}
	H = \frac{1}{2} \int \left(
	J^{- 1} \vec{\Pi} J \vec{\Pi} + \vB^2
	\right) + H_{\mathrm{C}} \, .
	\ee
	Here $\Pi = \delta / i \delta \vA$ is the momentum operator which represents the color electric field and $\vB$ is the color magnetic field. The magnetic term $\vB^2$ serves as a potential for the gauge field. Furthermore, $J = \operatorname{Det}(- \hat{D} \cdot \vec{\partial})$ is the Faddeev--Popov determinant, with $\skew2\hat{\vD} = \vec{\partial} + g \skew2\hat{\vA}$, $\skew2\hat{A}^{ab} = f^{acb} A^c$ being the covariant derivative in the adjoint representation of the gauge group denoted by the ``hat''. Finally, 
	\be
	\label{G26}
	H_{\mathrm{C}} = \frac{1}{2} \int J^{- 1} \rho J (- \skew2\hat{\vD} \cdot \vec{\partial})^{- 1} (- \partial^2) (- \skew2\hat{\vD} \vec{\partial})^{- 1} \rho
	\ee
	is the so-called Coulomb term which arises from the longitudinal part of the kinetic energy of the gauge field. Here 
	\be
	\label{G27}
	\rho^a = - \skew2\hat{\vA}^{ab} \vec{\Pi}^b +  \rho^a_q \, , \qquad \rho^a_q = q^\dag t^a q
	\ee
	is the color charge  density which contains besides the color charge density of the quark fields also a gluonic term. Let us stress that in the gauge fixed Hamiltonian Eq.~(\ref{G25}) Gauss' law has been exactly resolved, so gauge invariance is fully taken into account (at least as long as the Hamiltonian is treated exactly). In the gauge fixed theory the Faddeev--Popov determinant $J$ enters also the integration measure in the scalar product of wave functionals
	\be
	\label{G28}
	\langle \phi | \dots | \psi \rangle = \int D A \, J  \, \phi^*[A] \dots \psi[A] ,
	\ee
	where the integration runs over the transversal gauge fields only and must, in principle, be restricted to the so-called fundamental modular region. 
	\bi
	
	\no
	The aim of the Hamiltonian approach is to solve the functional Schr\"odinger equation $H \psi [A] = E \psi [A]$ for the vacuum wave functional $\psi [A]$. In the Yang-Mills sector we use the ansatz \cite{Feuchter2004,Feuchter2005}
	\be
	\label{G29}
	\psi [A] = \frac{1}{\sqrt{J}} \exp \left[
	- \frac{1}{2} \int A \omega A
	\right] \, ,
	\ee
	where $\omega (\vx, \vy)$ is a variational kernel which is determined by minimizing the expectation value of the gauge fixed Hamiltonian (\ref{G25}). Its Fourier transform represents the gluon energy since
	\be
	\label{395-GX}
	\langle \psi | A_i (\vx) A_j (\vy) | \psi \rangle = \frac{1}{2} t_{ij} (x) \omega^{- 1} (\vx, \vy)
	\ee
	with $t_{ij} (x) = \delta_{ij} - \partial_i \partial_j / \partial^2$ being the transverse projector. 
	\begin{figure}
		\centering
		\includegraphics[width=7cm,clip]{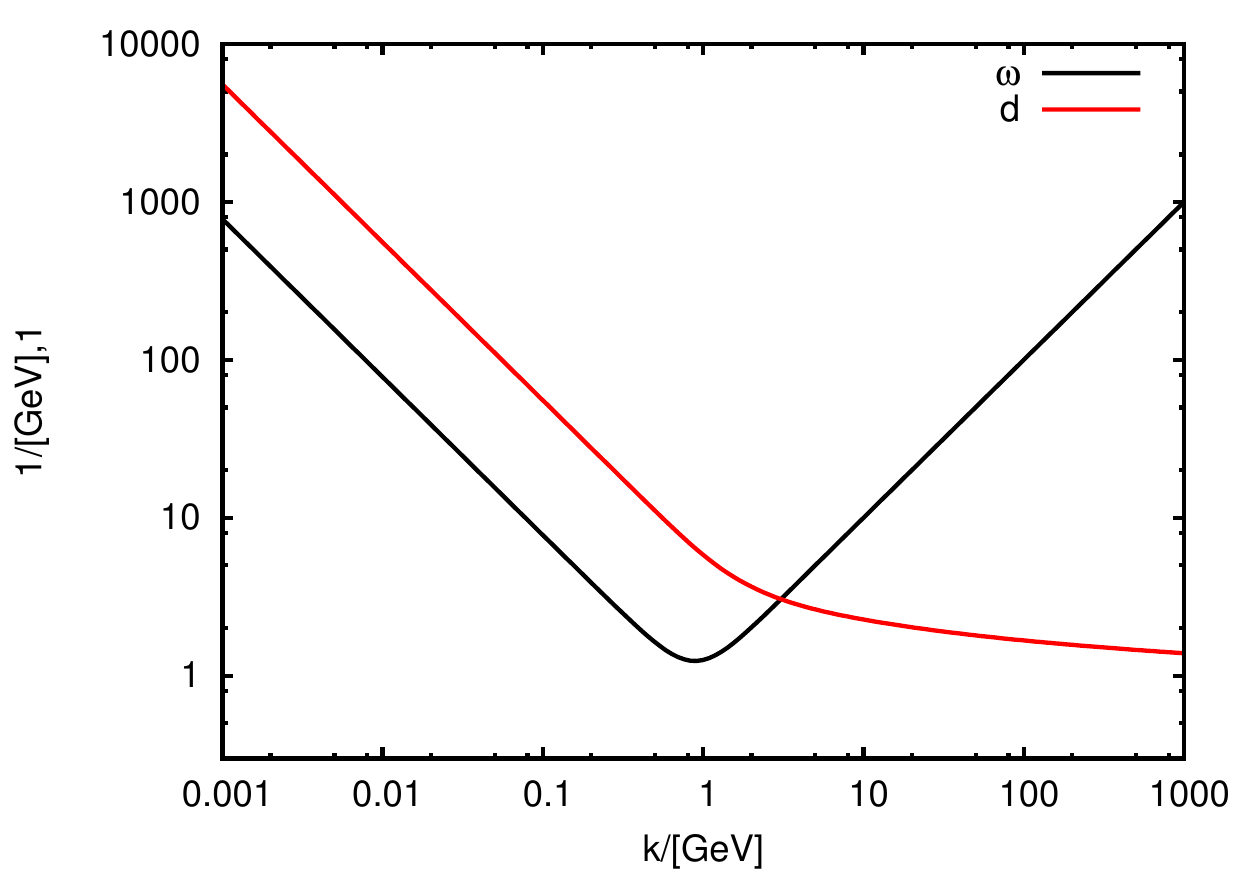}
		\caption{Numerical solution of the gluon energy (\ref{395-GX}) $\omega (k)$ and the ghost form factor (\ref{G32}).}
		\label{fig1} %
	\end{figure}%
\begin{figure}
	\centering
	\includegraphics[width=0.45\textwidth,clip]{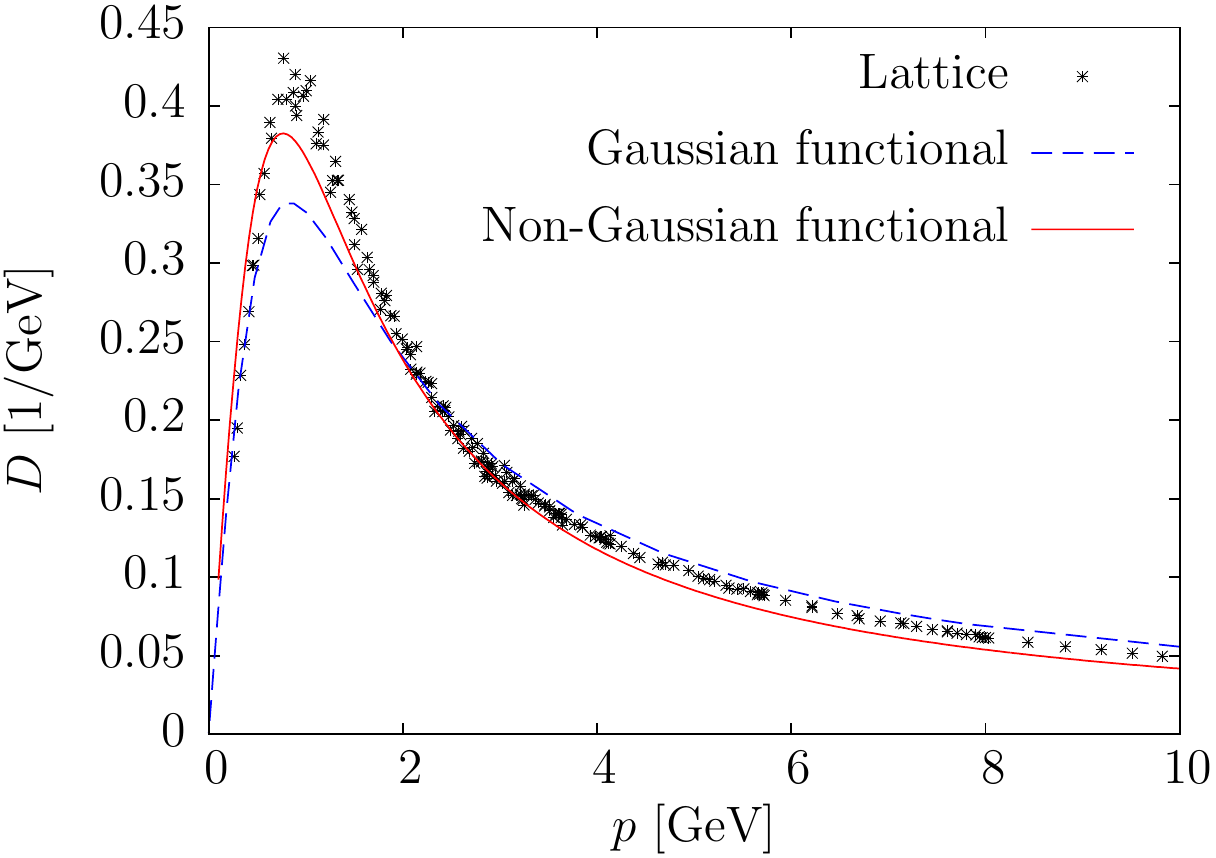}
	\caption{The static gluon propagator in Coulomb gauge calculated on the lattice for SU(2) gauge theory (crosses). The dashed and the full curves show the result of the variational 
		calculation using, respectively, a Gaussian and non-Gaussian ansatz for the vacuum wave functional.}
	\label{fig2}%
\end{figure}%
	 The resulting gluon gap equation can be solved analytically in the infrared and, furthermore, was solved numerically in the whole momentum regime. The result is shown in Fig.~1. The gluon energy $\omega (p)$ is infrared diverging signaling confinement while it behaves at large momenta like the photon energy in agreement with asymptotic freedom. Figure~\ref{fig2} shows the lattice data for the static gluon propagator (\ref{395-GX}) obtained in Ref.~\cite{BQR2009} together with  the result of the variational calculation both for the color group SU(2). One finds excellent agreement in the infrared and in the ultraviolet. However, in the mid-momentum regime some strength is missing. The agreement with the lattice is considerably improved when a non-Gaussian ansatz for the vacuum wave functional is used, which contains in the exponent besides the quadratic term also a cubic and quartic term, see Ref.~\cite{CR2010}. Note also that the lattice results can be perfectly fitted by the so-called Gribov formula \cite{Gribov1978,*Zwanziger:1988jt,*Zwanziger:1989mf,*Zwanziger:1992qr}
	\be
	\label{G31}
	\omega (\vp) = \sqrt{\vp^2 + M^4 / \vp^2} \, 
	\ee
	with a Gribov mass $M \simeq 880$\,MeV. 
	\bi
	
	\no
	A central role in Gribov's confinement scenario \cite{Gribov1978,*Zwanziger:1988jt,*Zwanziger:1989mf,*Zwanziger:1992qr} plays the ghost form factor $d (\vp)$ which is defined by
	\be
	\label{G32}
	\langle \psi | (- \skew2\hat{\vD} \vec{\partial})^{- 1} | \psi \rangle = d / (- \Delta) 
	\ee
	and embodies the non-Abelian features of Yang-Mills theory in the sense that it describes the deviation of the latter from electrodynamics where the ghost form factor is unity. It can be shown \cite{Reinhardt2008} that the inverse of this quantity represents the dielectric function of the Yang-Mills vacuum $\varepsilon (\vp) = d^{- 1} (\vp)$. By the so-called horizon condition $d^{- 1} (0) = 0$, which is fulfilled by the lattice data and which enters as a boundary condition in the variational calculation, the dielectric constant vanishes in the infrared making the Yang-Mills vacuum a perfect color dielectricum, which is nothing but a dual superconductor. In this way Gribov's confinement scenario is directly connected with the dual Meissner effect of the magnetic monopole picture of confinement. One can also show that the infrared divergence of the ghost form factor, i.e.~the horizon condition, $d^{- 1} (0) = 0$, disappears when one removes the so-called center vortices from the field configurations of the lattice functional integral, establishing also the connection of Gribov's scenario with the center vortex picture of confinement. 
	\bi
	
	\no
	In the quark contribution to the Coulomb term the Faddeev--Popov determinant drops out and the Yang-Mills vacuum expectation value of the Coulomb term gives rise to a potential acting between the static color charges 
	\be
	\label{G33}
	V (\vx, \vy) = \langle \psi | (- \skew2\hat{\vD} \cdot \vec{\partial})^{- 1} (- \partial^2) (- \skew2\hat{\vD} \cdot \vec{\partial})^{- 1} | \psi \rangle \, .
	\ee
	\begin{figure}
		\centering
					\includegraphics[width=7cm,clip]{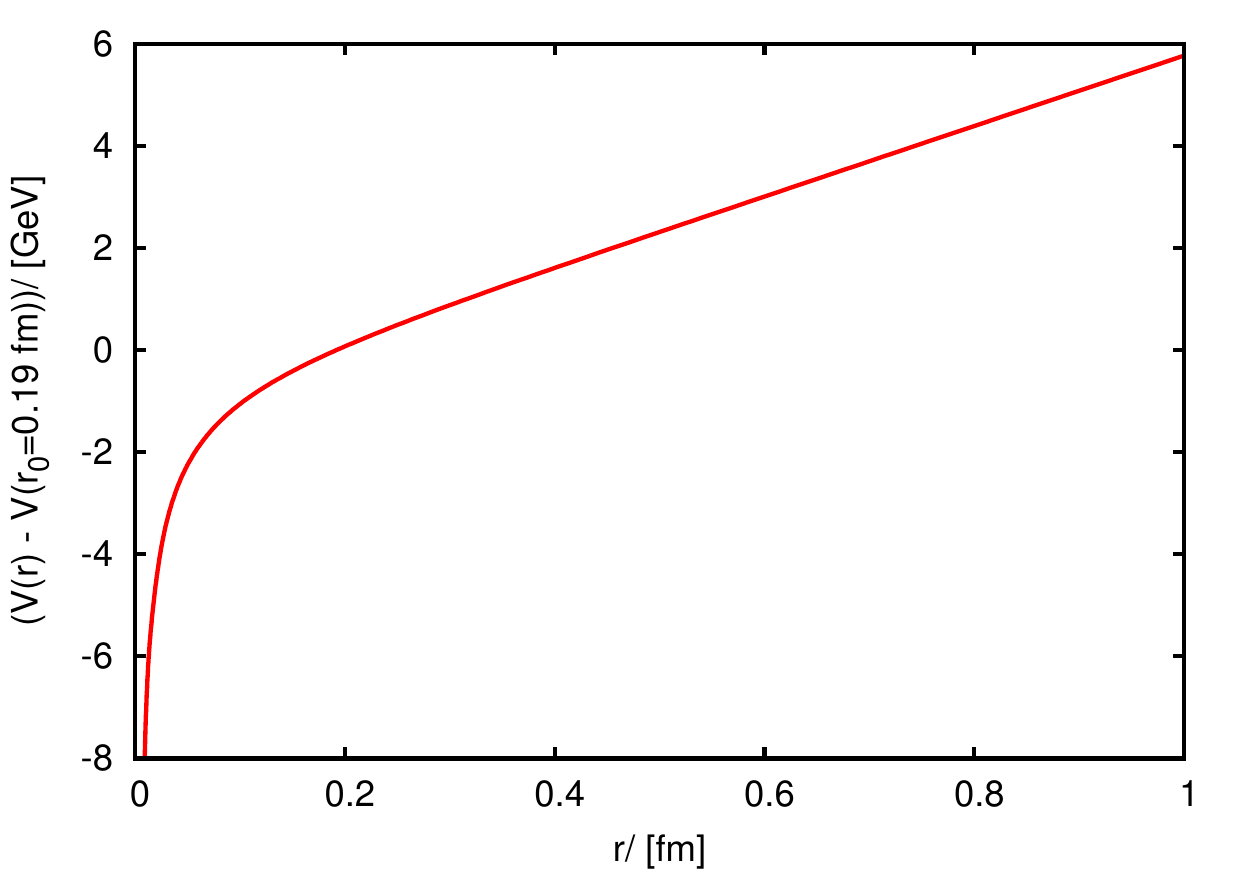}
						\caption{Non-Abelian Coulomb potential (\ref{G33}) obtained within the variational approach \cite{ERS2007}. }
		\label{fig3}%
	\end{figure}%
	This potential shown in Fig.~\ref{fig3}. It behaves at small distances like the ordinary Coulomb potential but increases at large distances linearly with a coefficient $\sigma_{\mathrm{C}}$, referred to as Coulomb string tension. It was shown that the Coulomb string tension represents an upper bound to the Wilsonian string tension \cite{Zwanziger2003} and, furthermore, that it is related not to the temporal but to the spatial string tension \cite{BQRV2015}. Like the latter it increases with the temperature above the deconfinement phase transition.
	\bi
	
	\no
	\subsection{Quark sector}\label{sect3.2}
	
\begin{figure}
		\centering
					\includegraphics[width= 7cm,clip]{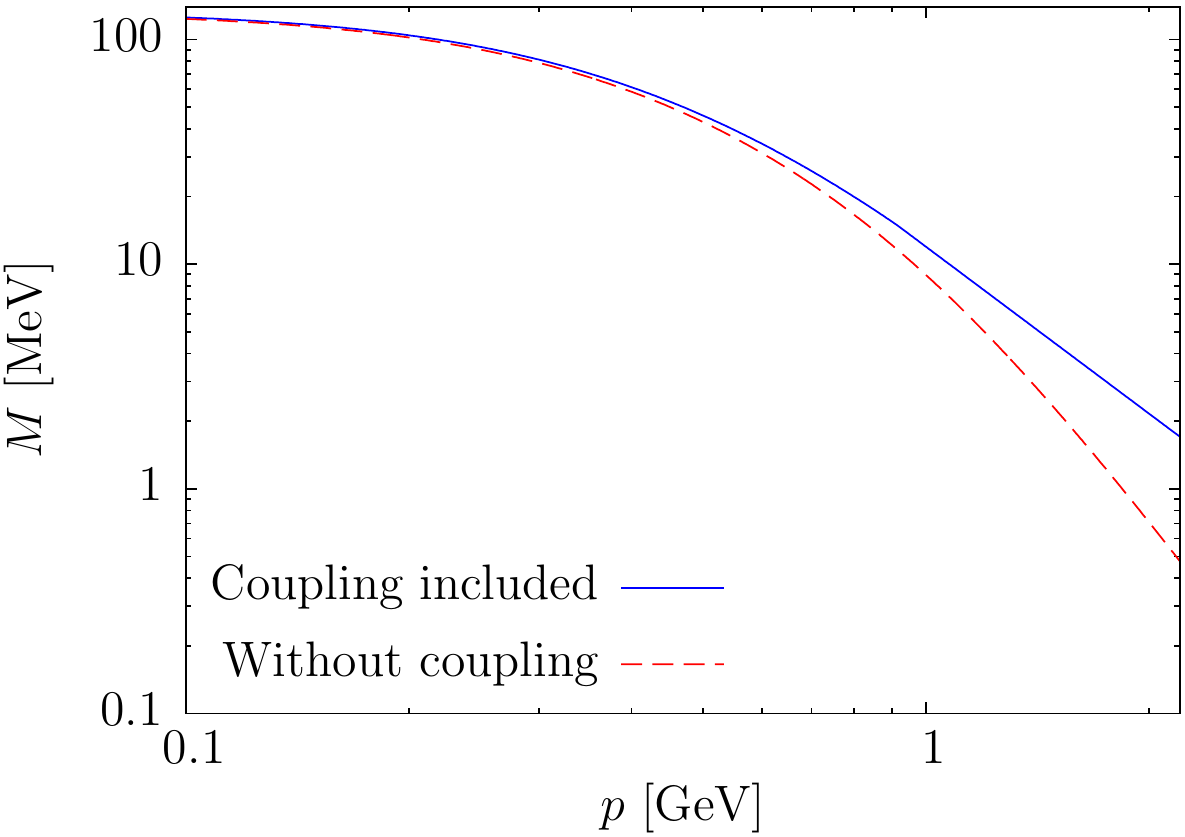}
					\caption{Mass function obtained from the (quenched) solution of the quark gap equation. Results are presented for $g \simeq 2.1$ (full curve) and $g = 0$ (dashed curve).}
		\label{fig4}%
	\end{figure}

\begin{figure}
	\centering
	\includegraphics[width=7cm,clip]{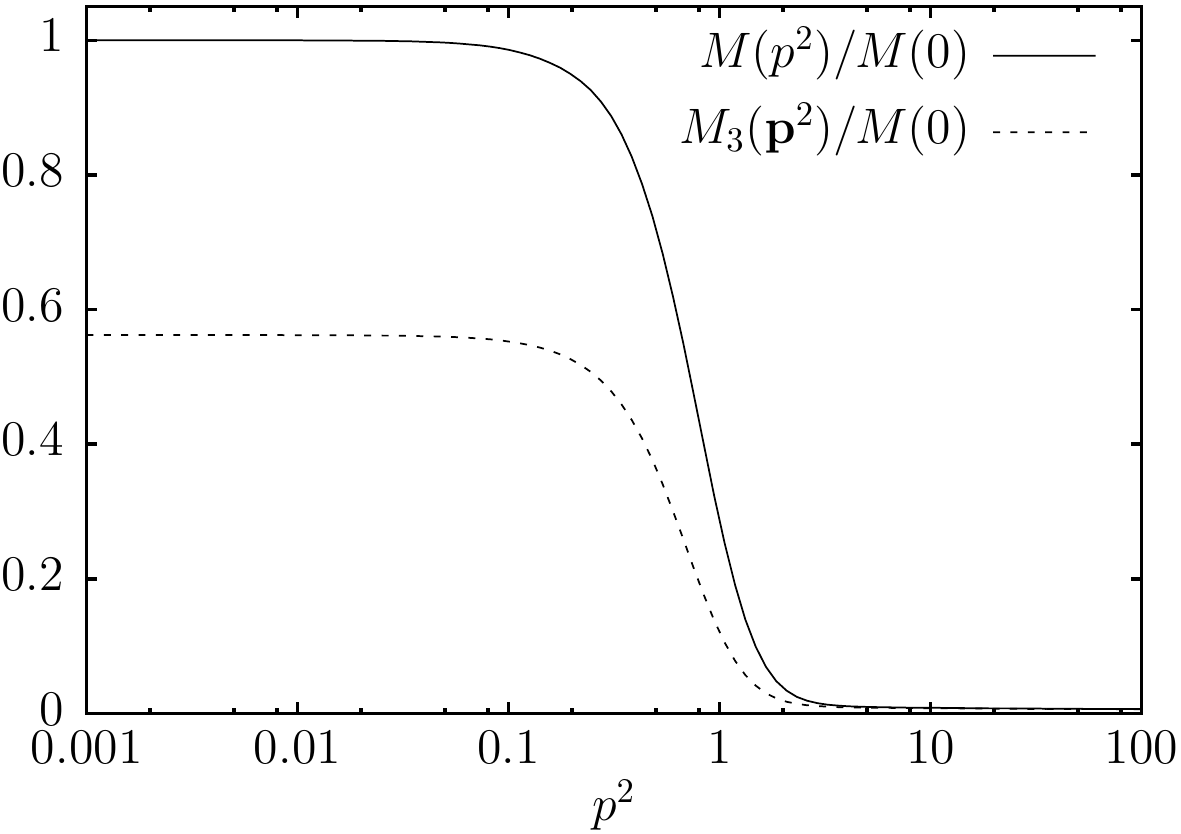}
	\caption{Comparison between the full mass function $M(p^2)$ in Landau gauge (continuous line) and the mass function $M_3(\mathbf{p}^2)$ of the equal-time propagator (dashed line), see ref. \cite{Campagnari:2018flz}.}
	\label{fig5}%
\end{figure}

	When the quarks are included one has to keep the quark part $\rho_q$ in the color charge density (\ref{G27}) and add the Dirac Hamiltonian 
	\be
	\label{G34}
	H_q = \int d^3 x \, q^\dag (x) \left[
	\vec{\alpha} (\vp + g \vA (x)) + \beta m_0
	\right] q (x) \, ,
	\ee
	where $\vec{\alpha}, \beta$ are the usual Dirac matrices and $m_0$ denotes the current mass of the quarks, which we will ignore in the following. For the vacuum wave functional of the quarks the following ansatz 
	\be
	\label{G35}
	\langle A | \phi \rangle = \exp \biggl\{ \int q^\dagger_+ \Bigl[ s \beta + \bigl(v + w \beta\bigr) \vec{\alpha} \cdot \vA \Bigr] q_- \biggr\}
	\ee
	was used, where $q_\pm$ denotes the positive and negative energy part of the quark field and $s$, $v$, $w$ are variational kernels. This ansatz contains explicitly the coupling of the quarks to the spatial gluons. When this coupling is neglected, $v = w = 0$, the ansatz (\ref{G35}) reduces to the BCS type wave functional considered in refs. \cite{FM1982,Adler1984,AA1988}. As we will see in  a moment the inclusion of the coupling of the quarks to the spatial gluons is absolutely necessary in order to reproduce the phenomenological value of the quark condensate as shown in Ref.~\cite{Pak2013} where the ansatz (\ref{G35}) with $w = 0$ was used. The advantage of keeping both Dirac structures in the quark gluon coupling in Eq.~(\ref{G35}) is that all UV divergences cancel in the resulting variational equations, see refs. \cite{QCDT0,QCDT0Rev}. The variational equations for $v$, $w$ can be explicitly solved in terms of the scalar kernel $s$ and the gluon energy $\omega$. For the quark sector one finds then a single non-linear equation which is conveniently expressed in terms of the effective quark mass 
	\be
	\label{36}
	M (\vp) = \frac{2 p s(\vp)}{1 - s^2(\vp)} .
	\ee
	In the numerical calculation we use for the static quark potential (\ref{G33})
	\be
	\label{4457-GX}
	V_{\mathrm{C}} (|\vp|) = \frac{g^2}{\vp^2} + \frac{8 \pi \sigma_{\mathrm{C}}}{(\vp^2)^2}
\ee
with the Coulomb string tension $\sigma_{\mathrm{C}}$ defining the scale of the theory. Furthermore the quark-gluon coupling constant $g$, which is the same as in eq. (\ref{G34}), was chosen to reproduce the phenomenological value of the quark condensate
	\be
	\label{G39}
	\langle \bar{q} q \rangle = (- 235\,\mathrm{MeV})^3 \, ,
	\ee
	which requires $g \simeq 2.1$ for the value $\sigma_{\mathrm{C}} = 2.5 \sigma$ of the Coulomb string tension favored by our lattice calculation \cite{BQR2009}. This corresponds to $\alpha = g^2 / 4 \pi \approx 0.35$ at the chiral symmetry breaking scale. The resulting effective quark mass is shown in Fig.~\ref{fig4} as function of the momentum. For sake of comparison we also show the mass function which is obtained when the coupling of the quarks to the spatial gluons as well as the perturbative part of the Coulomb potential (\ref{4457-GX}) is omitted $(g = 0)$. Both curves agree in the infrared which is due to the fact that this regime is exclusively determined by the Coulomb potential (\ref{4457-GX}). The two curves differ substantially only in the ultraviolet. This part gives, however, a significant contribution to the quark condensate. If the coupling of the quarks to the spatial gluons is omitted $(g = 0)$ one finds a quark condensate of 
	\be
	\label{G40}
	\langle \bar{q} q \rangle = (- 185\,\mathrm{MeV})^2 \, 
	\ee
	when the Coulomb string tension is chosen as before $(\sigma_{\mathrm{C}} = 2.5 \sigma)$. The infrared value of the effective quark mass $M (0) = 140$\,MeV seems to be substantially too small compared to the effective mass extracted from the quark propagator calculated in Landau gauge on the lattice and in Dyson--Schwinger calculations. However, it was shown in Ref.~\cite{Campagnari:2018flz} that the effective quark mass extracted from the static quark propagator is significantly smaller than the one obtained from the 4-dimensional propagator, see Fig.~\ref{fig5}. 
	\bi
	
	\no
	The results obtained so far at zero temperature within the Hamiltonian approach to QCD in Coulomb gauge are rather encouraging to warrant also the extension to finite temperatures in the novel approach presented in section \ref{sec-2}. 
	\bi
	
	\no
	\section{QCD at finite temperatures}
	
	The extension of the above presented Hamiltonian approach to QCD in Coulomb gauge to finite temperatures is straightforward but numerically much more expensive since a single integral equation from the approach at zero temperature becomes a coupled set of integral equations at finite temperature, each included Matsubara frequency gives rise to an equation. In order to reduce the numerical expense in the following we will consider only the quark sector where we neglect the coupling of the quarks to the spatial gluons. In the approximation considered in section \ref{sect3.2} the quark sector then decouples from the Yang-Mills sector. The influence of the Yang-Mills sector is then completely contained in the non-abelian Coulomb potential (\ref{G33}) which, for simplicity, we will replace by its infrared limit given in Eq.~(\ref{4457-GX}) with $g = 0$. The quark Hamiltonian then reduces to 
\be
\label{G42}
H = \int \mathrm{d}^3 x \, q^\dagger (\vx) \vec{\alpha} \cdot \vp q (\vx)
    + \frac{1}{2} \int \mathrm{d}^3 x \int \mathrm{d}^3 y \, \rho (\vx) \, V_{\mathrm{C}} (\vx - \vy) \, \rho (\vy) ,
\qquad \rho^a (\vx) = q^\dag(\vx) t^a q(\vx) ,
\ee
	which, in fact, is the model considered in refs. \cite{FM1982,Adler1984}. The corresponding functional Schr\"odinger equation is then solved variationally with the ansatz (\ref{G35}) with $v = w = 0$, resulting in the quark gap equation 
	\be
	\label{G43}
	M (\vp) = \frac{C_F}{2} \il_L \frac{d^3 q}{(2 \pi)^3} V_{\mathrm{C}} (|\vp - \vq|) \frac{M (\vq) - M (\vp) \skew2\hat{\vp} \cdot \skew2\hat{\vq}}{\sqrt{\vq^2 + M (\vq)}} 
	\ee
	for the effective quark mass (\ref{36}) where the finite-temperature integration measure is defined in Eq.~(\ref{G11}) and we have used the shorthand notation $M (\vp) \equiv M (\vp_\perp, \omega_n)$.
	\bi
	
	\no
	We have solved the gap equation (\ref{G43}) in the Matsubara form (\ref{G11}) starting at high temperatures and in the Poisson resummed form (\ref{G14}) given by 
\be
\label{486}
M (\vp) = \frac{C_F}{2} \sum_{k=-\infty}^\infty (-1)^{k} \int \frac{d^3 q}{(2 \pi)^3} \, V_{\mathrm{C}}(\vq) \cos\bigl(\beta k(q_z+p_z)\bigr)
\, \frac{M(\vq) - \bigl[ 1+{\textstyle\frac{\vp\cdot\vq}{\vp^2}}\bigr] M(\vp)}{\sqrt{(\vq+\vp)^2 + M (\vq+\vp)}} \, .
\ee
	\begin{figure}[t!]
		\centering
		\includegraphics[width=0.45\linewidth]{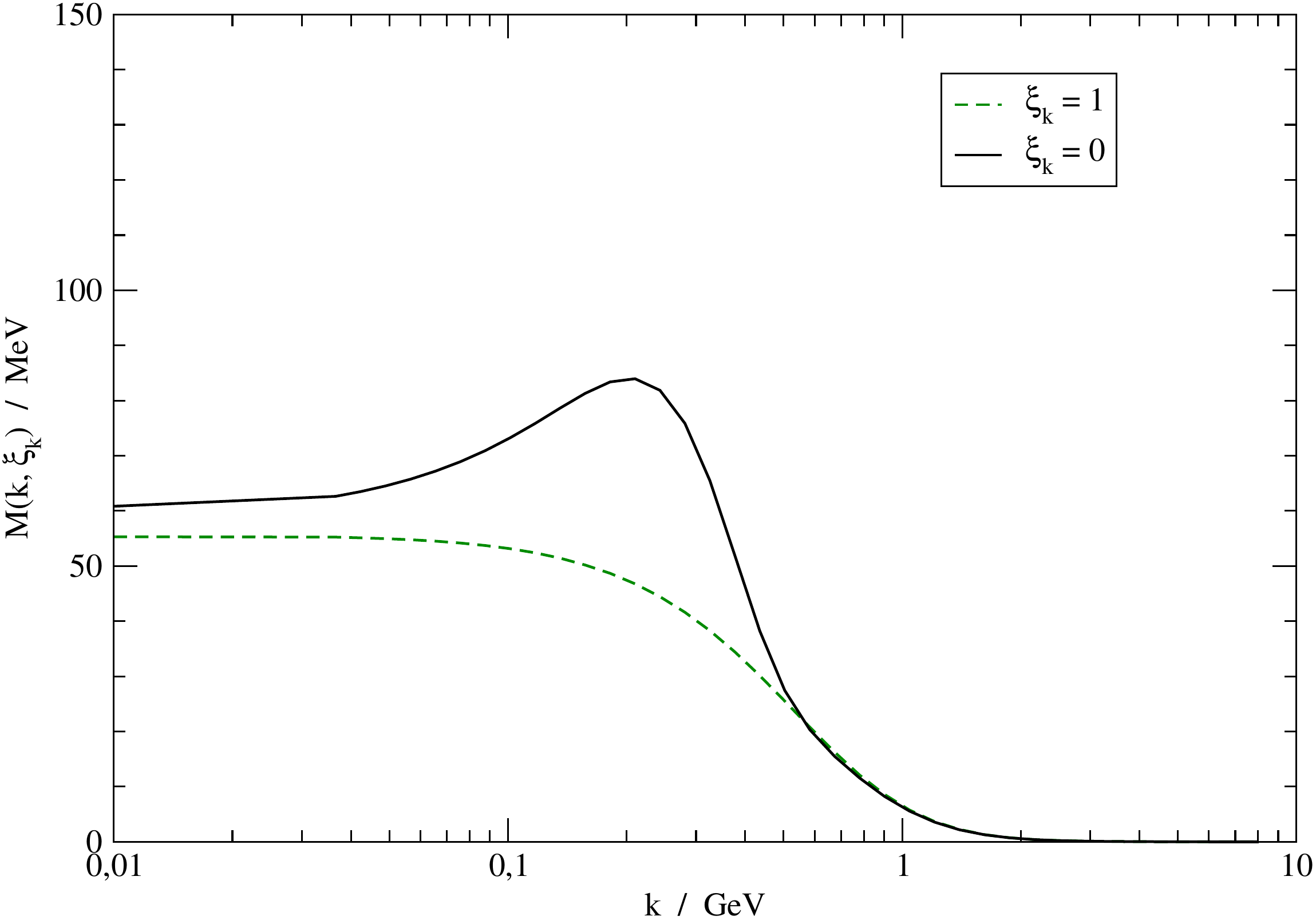}
		\qquad
		\includegraphics[width=0.45\linewidth]{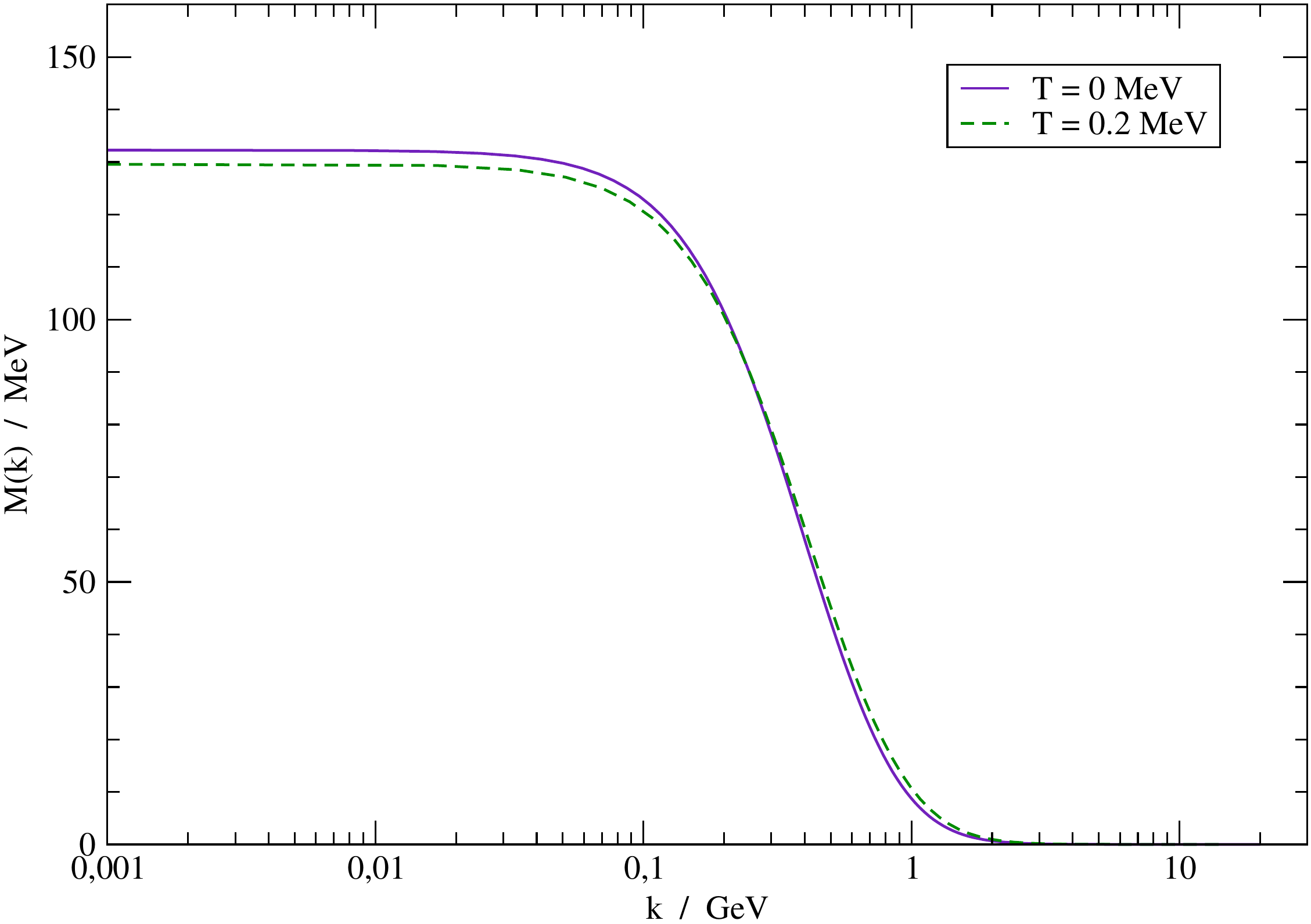}	
		\caption{\emph{Left:} Mass function $M(k,\xi_k)$ at $T=80\,\mathrm{MeV}$ 
			with the momentum $\vk$ pointing in various directions relative to the heat bath.
			\emph{Right:} Mass function $M(k,1)$ for small temperatures compared to the 
			$T=0$ limit.}
		\label{fig6}
	\end{figure}%
\begin{figure}[t!]
	\centering
	\includegraphics[width=0.8\linewidth]{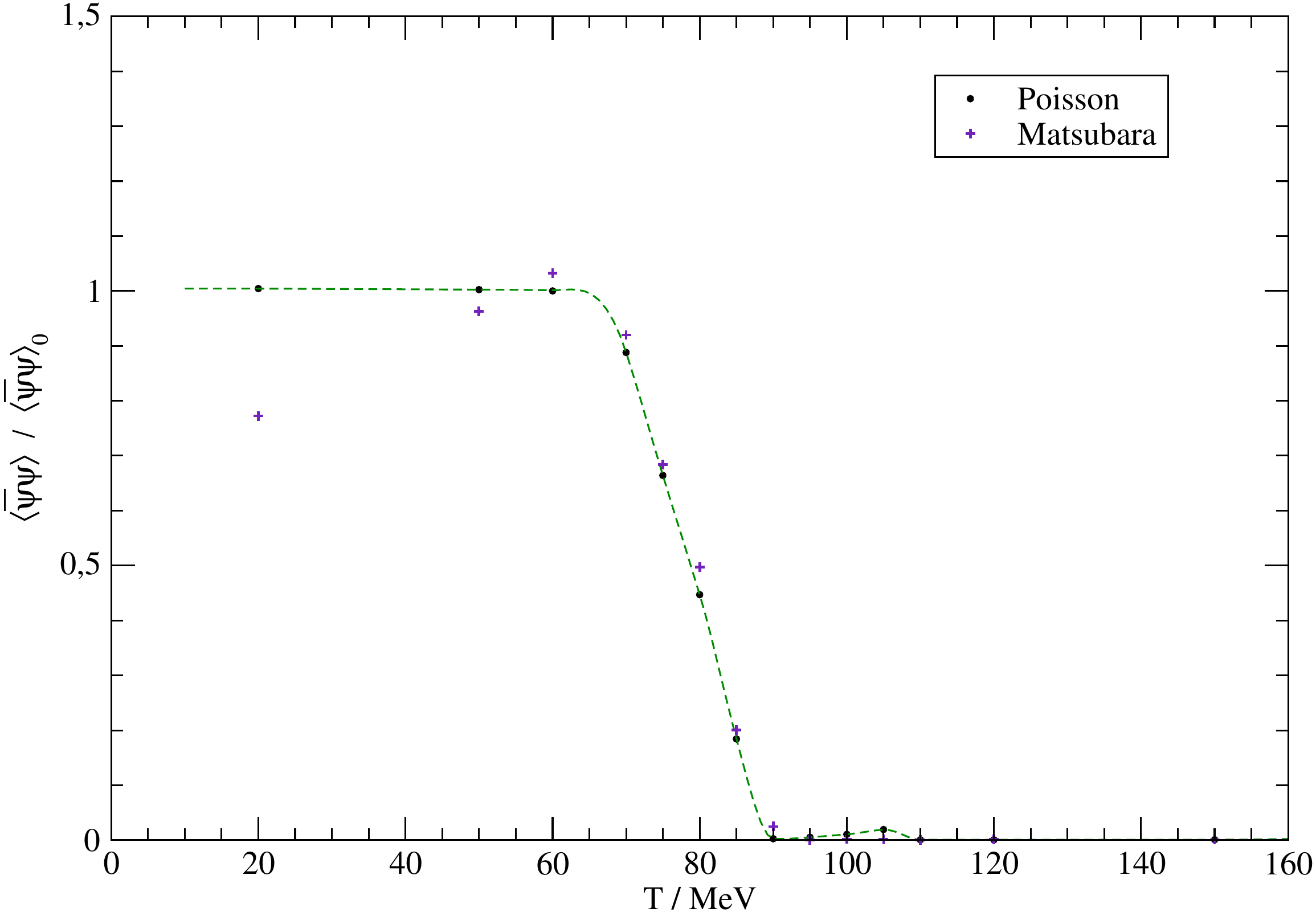}	
	\caption{Chiral condensate as a function of the temperature, from both 
		the Matsubara and Poisson formulation. The dashed line indicates a fit 
		to the Poisson data from which the critical temperature is determined.}
	\label{fig7}
\end{figure}%
	 This form is most convenient for small temperatures where only a few terms of the Poisson sum have to be included (remember the $k = 0$ term represents the zero-temperature contribution). The numerical expense is larger for the Poisson resummed form due to the oscillating behaviour of the integrand. However, it is advantageous in the sense that it is manifestly infrared finite. In the Matsubara form (\ref{G43}) a spurious infrared singularity appears when the sum over Matsubara frequencies is restricted to a finite number of terms. When the Matsubara sum is carried out to infinite order (which effectively is done in the Poisson resummed form) the infrared singularity disappears. Therefore the numerical solution of the gap equation in the Matsubara frequency representation has to be done with great care. We solve the Poisson resummed form starting from zero temperature until above the chiral phase transition regime and solve the equation (\ref{G43}) in the Matsubara representation at large to moderate temperatures below the deconfinement phase transition \cite{Quandt:2018bbu}. 
	 There is a broad overlap regime where both representations can be used and yield the same result. Figure~\ref{fig6} shows the resulting effective quark mass at finite temperatures for two extreme cases of the orientation of the external momentum relative to the compactified dimension. Compared to the zero-temperature solution there is a substantial reduction in the infrared mass already well below the temperatures where the phase transition occurs. This reduction of the effective quark mass has, however, little effect on the quark condensate until near the phase transition regime. This can be seen from Fig.~\ref{fig7} where the quark condensate is shown as function of the temperature. The figure contains both the results from the Matsubara representation as well as from the Poisson resummed form of the gap equation. Both results agree very well in a broad transition regime where the chiral phase transition occurs. This transition is seen to be of second order. From the numerical data one extracts a critical temperature of 
	\be
	\label{G45}
	T_\chi = 0.13 \sqrt{\sigma_{\mathrm{C}}} \, .
\ee
Using for the Coulomb string tension the value favoured by our lattice calculations $\sigma_{\mathrm{C}} = 2.5 \sigma$ one finds a critical temperature of $T_\chi \simeq 92$\,MeV. This temperature is too small compared to the lattice result $T_\chi^{lat} = 155$\,MeV. A smaller critical temperature is expected from the present calculations since we have neglected the coupling of the quarks to the spatial gluons. As we have seen in section \ref{sect3.2} this coupling increases the value of the chiral quark condensate. If we choose the Coulomb string tension to reproduce the phenomenological value of the quark condensate (\ref{G39}), which requires $\sigma_{\mathrm{C}} = 4.1 \sigma$, we find indeed a substantial larger critical temperature of $T_\chi = 115$\,MeV which, however, is still too small compared to the lattice result. One should notice, however, that in the Adler-Davis model considered here the temperature effects of the gluon sector are completely ignored. In particular, the Coulomb string tension is known to increase with the temperature \cite{BQRV2015} resulting via Eq.~(\ref{G45}) in an increase in the critical temperature of the chiral phase transition. Furthermore, in the present calculations we have neglected the ultraviolet part of the non-Abelian Coulomb potential whose effect on the critical temperature is, however, difficult to estimate. 
\bi

\no
\begin{figure}[t!]
	\centering
	\includegraphics[width=0.8\linewidth]{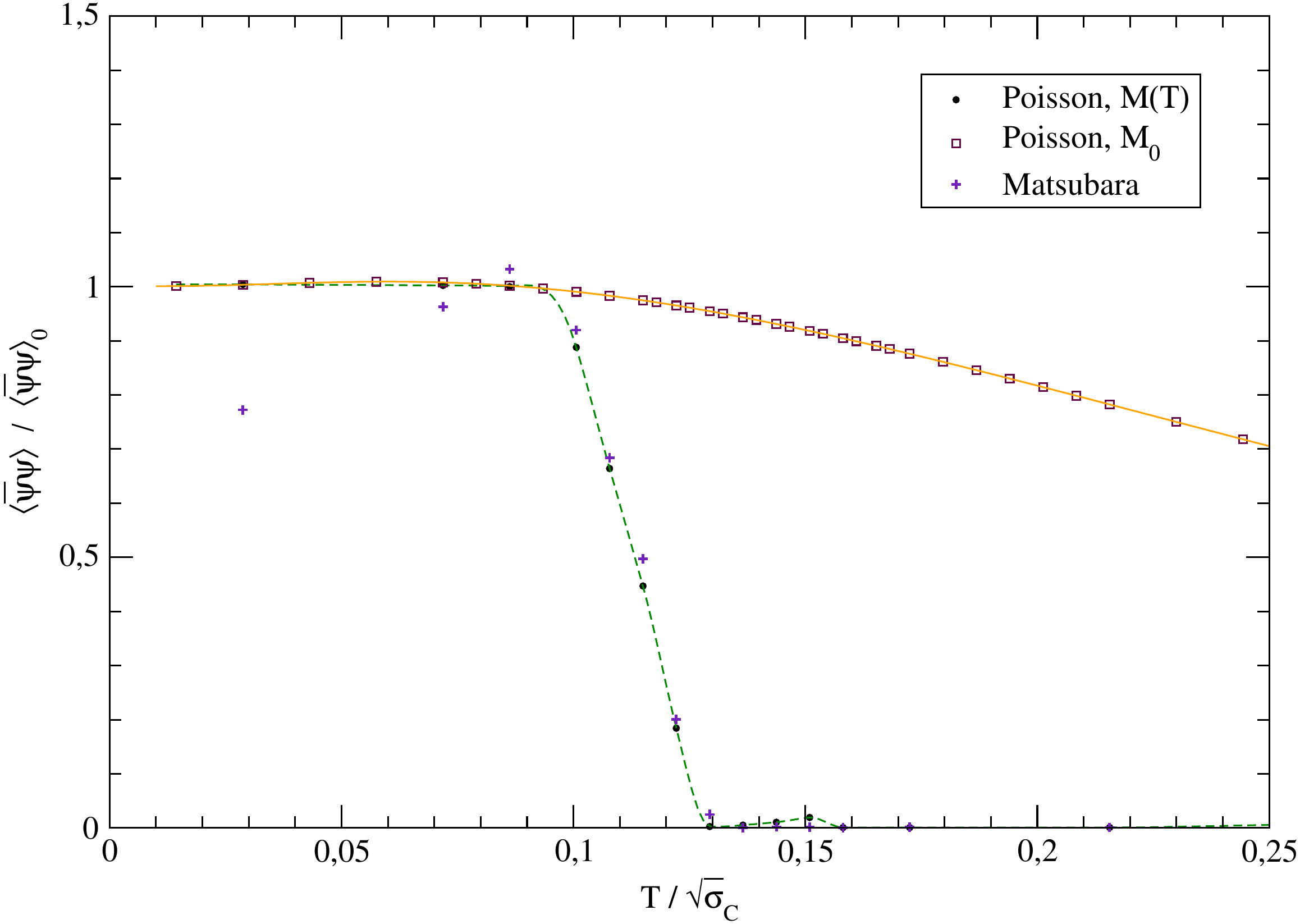}	
	\caption{The quark condensate as function of the temperature calculated from the zero- and finite-temperature solutions of the quark gap equation.}
	\label{fig8}
\end{figure}%
Let us also mention that in the canonical finite-temperature Hamiltonian approach where a quasi-particle ansatz is assumed for the Hamiltonian in the grand canonical density operator $\exp (- H/T)$ one finds an even smaller critical temperature of $T_\chi = 0.091 \sqrt{\sigma_{\mathrm{C}}}$. Finally Fig.~\ref{fig8} shows the quark condensate as function of the temperature when one uses the zero-temperature solution of the quark gap equation. As one can see this approximation is valid below the phase transition regime but fails completely as the temperature approaches the transition regime. The chiral condensate calculated with the zero-temperature solution decreases only slowly with the temperature and does not show a second order phase transition.
\bi

\no
The novel approach to finite temperatures within the Hamiltonian formulation assumes O(4) invariance of the Euclidean field theory. This implies that all components of the gauge field are treated on an equal footing. This is certainly not the case for
the Adler-Davis model. The potential term arises from the correlator of the temporal gluons while the spatial gluons are neglected in this model.
The present approach to finite-temperature Hamiltonian quantum field theory, which assumes O(4) invariance including covariance in the treatment of the gauge fields, when applied to the Adler-Davis model in fact yields the thermodynamics of the covariant extension of this model. So the present approach is not only superior in the sense that it does not require additional approximation to the grand canonical density operator. It also yields automatically the finite-temperature theory of the covariant extension of a non-covariant model field theory.
In general, the present approach when applied to a non-relativistic invariant theory (i.e.~non O(4) invariant theory in Euclidean space) yields the finite-temperature theory of a relativistic extension.

\section{Conclusion}

In this talk I have presented a novel approach to the Hamiltonian formulation of quantum field theory at finite temperatures by compactifying a spatial dimension. I have shown that this approach agrees with the results of the traditional canonical approach when exact (analytic) calculations are possible. The novel approach is advantageous over the traditional one since it does not require an explicit calculation of the trace of the grand canonical density operator, which in an interacting quantum field theory is difficult to handle and necessitates further approximations. Rather, in the novel approach the complete finite-temperature theory is encoded in the vacuum state on the spatial manifold $S^1 (L) \times \RR^2$ with one dimension compactified to a circle $S^1 (L)$. The circumference of the circle represents the inverse temperature. In Ref.~\cite{RH2013} this novel approach has been used to calculate the effective potential of the Polyakov loop in pure Yang-Mills theory using, however, the zero-temperature gluon and ghost propagator. The correct order of the deconfinement phase transition (second order for SU(2) and first order for SU(3)) were obtained with critical temperatures in the range between 270\,MeV and 290\,MeV. Unfortunately these calculations were done using the zero-temperature variational solution of the Yang-Mills Schr\"odinger equation. These calculations should be repeated using the finite-temperature variational solutions. At the moment we calculate the effective potential of the Polyakov loop including the quark Coulomb term which is crucial in the quark sector as we have seen in the study of the chiral phase transition. When the infrared part of the non-Abelian Coulomb potential (\ref{4457-GX}) is neglected one does not find spontaneous breaking of chiral symmetry for reasonable values of the coupling of the quarks to the spatial gluons. Therefore this term is absolutely necessary for the description of the infrared behaviour of the quark sector and should hence also be included in the calculation of the effective potential of the Polyakov loop, the order parameter of confinement.
Finally the present approach should be extended to finite quark chemical potential which is the regime of most interest in the QCD phase diagram. 

\section*{Acknowledgment}
This work was supported in part by the Deutsche Forschungsgemeinschaft under
contracts DFG-Re856/9-2 and Re856/10-1.

\bibliography{QCDT0_nizza}
%\begin{thebibliography}{}
%
% and use \bibitem to create references.
%

%\end{thebibliography}

\end{document}